\documentclass[11pt,modern]{aastex} 
\usepackage{hyperref}
\usepackage{natbib}
\usepackage[section]{placeins}

\newcommand{\BL}{{\textit{Breakthrough Listen }}}
\newcommand{\BLp}{{\textit{Breakthrough Listen}}} 

\slugcomment{To be Published in PASP}
\received{2016 Dec 6}
\accepted{2017 Jan 4}

\shorttitle{Breakthrough Listen Target Lists}
\shortauthors{Isaacson et al.}

\begin{document}


\title{The Breakthrough Listen Search for Intelligent Life: Target Selection of Nearby Stars and Galaxies}


\author{Howard Isaacson\altaffilmark{1,2}, Andrew P. V. Siemion\altaffilmark{1,3,4,5}, Geoffrey W. Marcy\altaffilmark{6}, Matt Lebofsky\altaffilmark{1}, Danny C. Price\altaffilmark{1}, David MacMahon\altaffilmark{1}, Steve Croft\altaffilmark{1}, David DeBoer\altaffilmark{1}, Jack Hickish\altaffilmark{1}, Dan Werthimer\altaffilmark{1}, Sofia Sheikh\altaffilmark{1}, Greg Hellbourg\altaffilmark{1}, J. Emilio Enriquez\altaffilmark{1}}

\email{hisaacson@berkeley.edu}

\altaffiltext{1}{Astronomy Department, University of California,
    Berkeley, CA, USA}
\altaffiltext{2}{corresponding author}
\altaffiltext{3}{ASTRON, Netherlands Institute for Radio Astronomy, Dwingeloo, NL}

\altaffiltext{4}{Radboud University, Nijmegen, NL}

\altaffiltext{5}{SETI Institute, Mountain View, California, USA}
\altaffiltext{6}{Astronomy Department, University of California,
    Berkeley, CA, USA Professor Emeritus}

\begin{abstract}

We present the target selection for the Breakthrough Listen search for extraterrestrial intelligence during the first year of observations at the Green Bank Telescope, Parkes Telescope and Automated Planet Finder. On the way to observing 1,000,000 nearby stars in search of technological signals, we present three main sets of objects we plan to observe in addition to a smaller sample of exotica. We choose the 60 nearest stars, all within 5.1 pc from the sun. Such nearby stars offer the potential to observe faint radio signals from transmitters having a power similar to those on Earth. We add a list of 1649 stars drawn from the Hipparcos catalog that span the Hertzprung-Russell diagram, including all spectral types along the main sequence, subgiants, and giant stars. This sample offers diversity and inclusion of all stellar types, but with thoughtful limits and due attention to main sequence stars. Our targets also include 123 nearby galaxies composed of a ``morphological-type-complete'' sample of the nearest spirals, ellipticals, dwarf spherioidals, and irregulars. While their great distances hamper the detection of technological electromagnetic radiation, galaxies offer the opportunity to observe billions of stars simultaneously and to sample the bright end of the technological luminosity function.  We will also use the Green Bank and Parkes telescopes to survey the plane and central bulge of the Milky Way. Finally, the complete target list includes several classes of exotica, including white dwarfs, brown dwarfs, black holes, neutron stars, and asteroids in our Solar System.
\end{abstract}



\section{Introduction}

For the last century, searches for extraterrestrial intelligence (SETI) have been conducted by gathering and analyzing electromagnetic radiation, arriving from beyond Earth, and filtering for characteristics not immediately explainable by naturally occurring processes in the universe, \citep{Cocconi1959}, \citep{Drake1961}. While the first SETI searches focused on radio wavelengths, more recently optical and infrared searches have also been undertaken. Aside from direct detection of electromagnetic radiation, in SETI one can also pursue the detection of an altered environment that is only explained by the presence of advanced technology, such as detection of Dyson spheres \citep{WrightJ2014a}, or the presence of artificially created molecules such as CFCs in an exoplanet atmosphere \citep{Lin2014}. 

Radio SETI searches take advantage of the transparency of the Earth's atmosphere at radio wavelengths and the ability of radio waves to pass through the interstellar medium with only small extinction \citep{Tarter2001, Werthimer2001, Tarter2011, Korpela2011, Siemion2013, Harp2016}. From the perspective of Earth technology, these searches would be sensitive to powerful radar such as the Arecibo Planetary Radar used to search for and study near Earth asteroids. 

Alien civilizations that are interested in sending high volumes of data with low energy may use optical lasers as a means of communication. However, such lasers are more prone to interstellar extinction, leading to a smaller search volume accessible in the optical spectrum. Previous searches in the optical have searched for laser emission \citep{WrightS2001, Reines2002, Howard2004, Stone2005, Tellis2015, Lacki2016, Howard2007}, as well as anomalous features in broadband photometry such as that available from the Kepler Space Mission  \citep{Walkowicz2014} that could point to the alteration of an alien environment such as Dyson spheres \citep{WrightJ2016}. 

SETI observations can be conducted by targeting large telescopes as a primary observer, or in a commensal or piggy-back fashion.  In the latter case, one performs an analysis on a duplicate or ancillary data stream produced during what are typically non-SETI observations \citep{Bowyer1983}.  This technique has the advantage that large amounts of the time on many telescopes can be utilized for SETI searches.  Radio telescopes can operate in this mode with zero loss of sensitivity for the primary observer. However, the secondary or commensal observer usually cannot control the field of view, the exposure times, nor receiver or filter bands of the light waves being observed.  While some telescopes, notably interferometric radio telescopes, can overcome these limitations significantly by presenting low level data products from individual antennas with wide fields of view, many observation parameters remain constrained. 

In targeted SETI searches, one carefully selects the directions in the sky that the telescope will be pointed.  One also selects receivers or filters, the spectral resolution, and the duration of the exposures. Targeted searches offer the opportunity to observe a specified population of targets, at well defined frequencies, spectral resolution, and flux limits. The resulting non-detections, or detections, can then be more easily translated into physical and statistical interpretations. Such observational specificity corresponds to well-defined properties, or upper limits, of  extraterrestrial civilizations such as their broadcasting power, frequency of occurrence near stars or galaxies, their number density in the universe, and the fraction of space occupied by their broadcast beams.  Flux thresholds can also be compared to the electromagnetic luminosity of the Earth at different frequencies. 

Recent targeted SETI searches have been carried out by \cite{Tarter2001}, \cite{Reines2002},  \cite{Howard2007}, \cite{Harp2016}, \cite{Siemion2013}, \cite{Maire2014}, \cite{WrightJ2014a}, \cite{WrightS2014}, and \cite{Tellis2015}.   Project Phoenix was arguably the world’s most comprehensive targeted search for extraterrestrial intelligence at radio wavelengths. It used the Parkes 64 m telescope for 16 weeks observing 200 stars, the Green Bank 140-foot telescope from 1996 to 1998, and finished by using the Arecibo 300 m radio telescope.  Phoenix targeted nearby, Sun-like stars based on the hypothesis that they represented plausible harbors of intelligent life. In total, Project Phoenix surveyed 800 stars within 65 parsecs, between frequencies of 1.0 to 3.0 GHz at 1 Hz-wide resolution. No signals were reported \citep{Backus2002}.


As a complement to targeted searches, wide-field surveys serve as a way to sample 10${^6}$ - 10${^8}$ more stars than pointed surveys, albeit at lesser sensitivity. \cite{Shostak1996} surveyed the Small Magellanic Cloud for narrow band signals using Project Phoenix hardware. \BL will use improved digital electronics and multi-beam receivers will allow for unprecedented sensitivity to a variety of signal types, including narrow band sources in the Galactic Plane and spatially extended nearby galaxies, such as Andromeda and the Large and Small Magellanic Clouds.

 The next-generation SETI program, \BLp, is initially utilizing three major telescopes, the Green Bank radio telescope, the Parkes radio telescope, and the Automated Planet Finder optical telescope and its Levy spectrometer at Lick Observatory. 
In this paper we present the method and results regarding the \BL selection of targets for these three telescopes.  The approach toward selecting targets is guided by four top level goals: 
1)  Sample all major types of stars over a range of masses, ages (including "evolved" stars), and elemental abundances; 2)  sample the region of the Milky Way Galaxy within 50 parsecs of the Sun, as it appears to be not substantially different from the rest of the disk of our Galaxy; 3) for galaxies, favor the nearest representatives of all of the major classes: elliptical, spirals, dwarf spheroidals, and irregulars. Radio observations of galaxies encompass tens of billions of stars in each telescope beam, allowing us to sample the rare but bright end of the extraterrestrial luminosity function; and 4) target some objects that, while defying the expected habitats for intelligent life due to seemingly harsh environments for life as we know it, are astrophysically anomalous or otherwise present intriguing underexplored possibilities for advanced life.

\section{Facilities}
\BL is currently using the Green Bank Telescope in West Virginia, the Automated Planet Finder at Lick Observatory and the Parkes Telescope in Australia, with plans to incorporate other large telescopes around the world. This complement of telescopes will allow wavelength coverage from 350 MHz -100 GHz in the radio and 374  to 950 nm in the optical. The combined receiver suite and instrumentation on these facilities allow for a wide array of observing possibilities.

The 100-m Green Bank Telescope, with newly installed \BL instrumentation, is currently capable of processing a 3.75 GHz bandwidth in dual polarization, and will soon expand to 10 GHz of bandwidth (MacMahon, in prep). All of the available receivers are capable of feeding the \BL instrument. Observations of the primary target list for \BL began on 2016 Jan 1, with observations of the first known exoplanet around a sun-like star, 51 Peg. Details of observations so far are listed in Section \ref{obs_to_date}.

\FloatBarrier

The 2.4-m Automated Planet Finder, with its Levy Spectrometer, is capable of producing high resolution optical spectra ($R=95,000$) over the wavelength range of 374  to 950 nm \citep{Radovan2014}. We use the telescope for 36 nights per year and search each spectrum for optical laser lines originating from the on sky area surrounding the target star on 36 nights per year. The decker size of $1\farcs0 \times 3\farcs0$ corresponds to 10 by 30 au for a star at 10 pc, and proportionally larger regions for more distant stars, making us sensitive to laser lines emitted from the entire planetary system around each target star as well as from foreground and background stars in the line of sight. The targeted signal-to-noise ratio is 100:1 per pixel with maximum exposure time of 20 minutes. Observations with the APF began in December 2015 and are detailed in Section \ref{obs_to_date}. 

\BL observations at the Parkes radio telescope in Australia began on 2016 Nov 7. The 64 m single dish telescope is currently capable of feeding the \BL instrument with a single pixel covering up to 1.25 GHz bandwidth at all of the available receivers, including the 13mm, H-OH, and MARS receivers. Installation of an ultra-wide-band feed that will cover 0.7-4 GHz along with an increase in bandwidth allowed by instrument upgrades will result in up to 5 GHz of total bandwidth in 2017 February, including observations using all 13 pixels of the Parkes Multibeam Receiver. The multi-beam and ultra-wide band feeds will be used for both a targeted star search and a survey of the Galactic plane and bulge which will provide a much shallower survey of billions of stars in the Milky Way. A special focus on the Galactic bulge will use the multi-bream receiver from 1.2 - 1.5 GHz. 

\section{The Stellar Sample}

The stellar sample is defined by two selection criteria. The first is a volume-limited sample of stars within 5 pc of the Sun. The second is a spectral class complete sample consisting of stars across the main sequence and some giant branch stars, all within 50 pc. Details of the selection criteria are described below. We combined the two sub-samples (5 pc and 5-50 pc) to produce the final set of 1709 target stars which are listed in Table \ref{table:stars_table}.

\subsection{The 5-Parsec Sub-sample}

We constructed a target sample containing all 60 known stars within 5 pc. Designed for observations with the GBT, APF, Parkes, and any other SETI observational projects such as NIROSETI \citep{WrightS2014}, this sub-sample contains all stars within 5 pc from the RECONS list \citep{Henry2006} and its online updates\footnote{ \url{http://www.recons.org}}, and the Gliese Catalog of Nearby Stars (3rd Edition) \\ \citep{Gliese1995}. We extracted their coordinates, trigonometric parallaxes, proper motions and photometry, including V and B-V magnitudes. Nearly all of these stars have large proper motions, in the range 0.1 - 5 arcsec $yr^{-1}$, accumulating during 16 years (since epoch 2000) to tens of arcseconds, comparable to the field of view of optical telescopes. We include proper motions for all observations.

\begin{figure}
\epsscale{.8}
\plotone{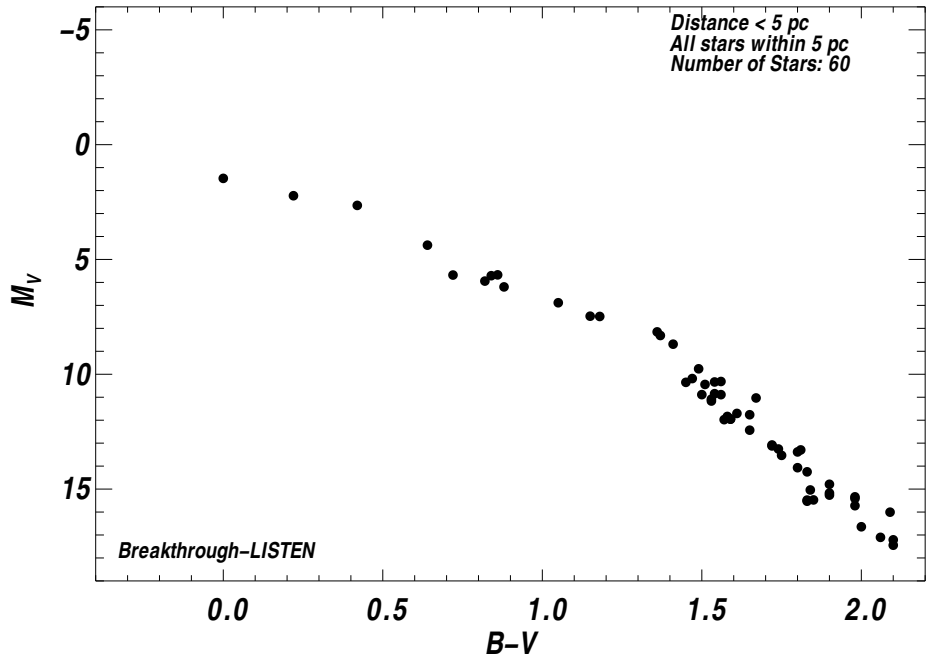}
\caption{ The H-R Diagram of the 5-parsec sample of target stars for \BLp. It includes all stars within 5 pc over the entire sky, north and south.
}
\label{fig1}
\end{figure}

Roughly half of the stars in the 5 pc sample have a binary companion. For any star having a  companion within an angular separation of 2 arcseconds, we simply enter the star in the \BL target list once, as they will not be spatially resolved at either optical or radio wavelengths. Binary stars with separations greater than 2 arcseconds receive an observation for each star, with knowledge that both stars may fall into the observable view of the telescope. The H-R Diagram of the 5-Parsec star sample is shown in Figure \ref{fig1}. Most of these stars are red, low mass, M dwarfs with B-V $>$ 1.0. The \BL observing strategy may in the future be modified to boost the exposure time of these nearby stars, to enable the detection of radio transmitters having modest (Earth-like) power measured in tens or hundreds of megawatts.

\subsection{5-50 Parsec Sub-Sample of Main Sequence and Giant Stars}

The second sub-sample of the \BL target list includes nearby main sequence and giant stars at all declinations, north and south, suitable for the Green Bank, APF, and Parkes telescopes. The selection criteria for the 5-50 pc target stars follow from the goal of searching nearby stars with a broad sampling of all types of main sequence stars. The 5-50 pc target stars are drawn from the Hipparcos Catalog \cite{Perryman1997}, which provides accurate coordinates, distances, and proper motions.  The following descriptions and figures illustrate the selection criteria of these 5-50 pc main sequence and giant stars.

\begin{figure}
\epsscale{.90}
\plotone{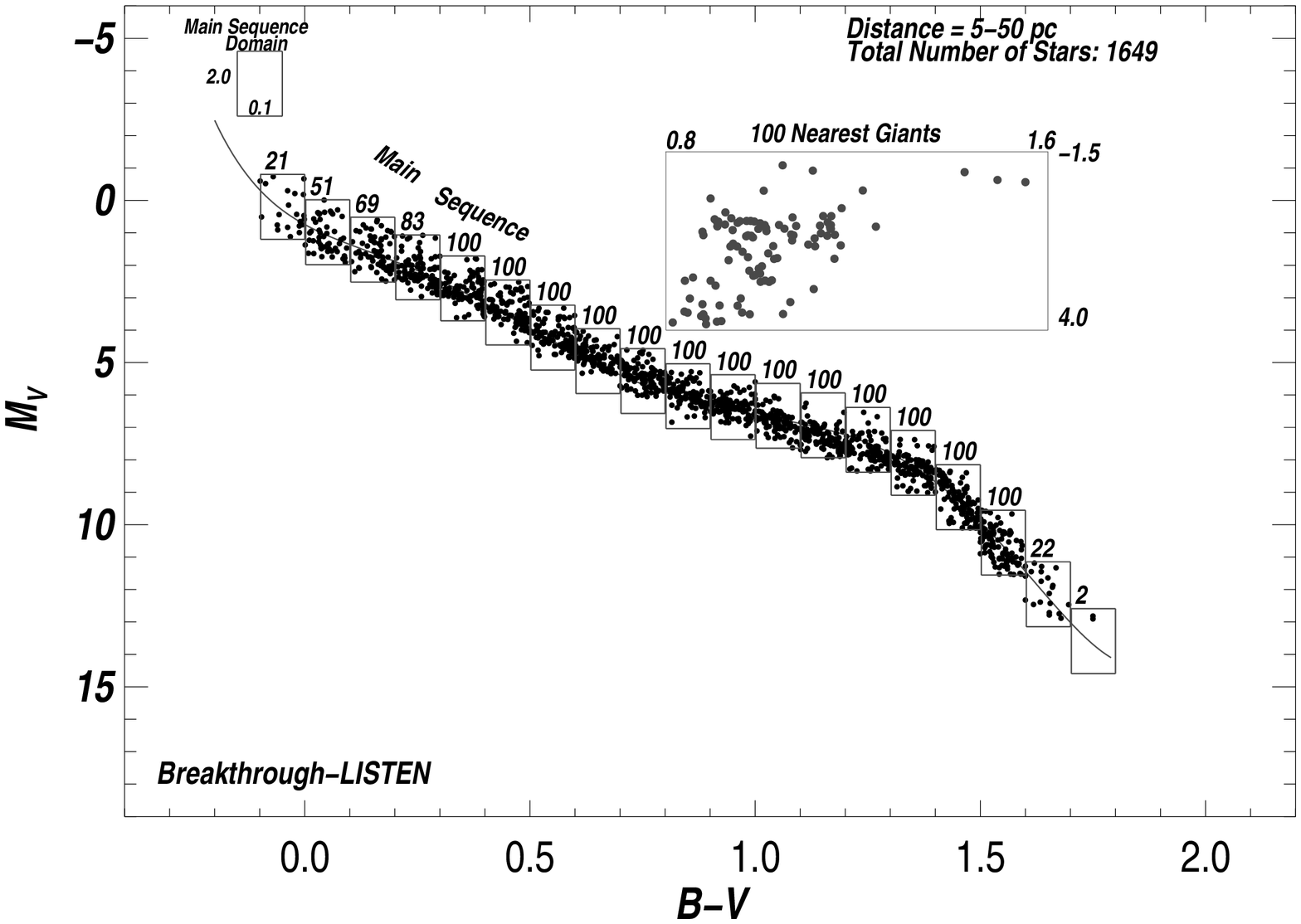}
\caption{The H-R Diagram of the 5-50 pc sub-sample of target stars for \BLp. Domains of $B-V$ and $M_V$ were constructed along the main sequence, within which the nearest 100 stars were selected to be included in the final \BL target list. In addition, a domain of giant stars was constructed and the nearest 100 such stars were identified. 
}
\label{fig2}
\end{figure}

Figure \ref{fig2} shows the H-R Diagram of the 5-50 pc sub-sample of stars. The figure exhibits rectangular domains along the Main Sequence stars in color (B-V) and visible luminosity ($M_{v}$) of size 0.1 mag and 2.0 mag, respectively, as shown in the upper left of Figure \ref{fig2}. Many domains contain more than 100 stars in this distance range. To spread more uniformly the distribution of target stars along the main sequence, the nearest 100 stars were identified and culled within each domain. The number of stars in each domain is indicated at the top of each domain. If there were fewer than 100 stars in a domain, all of the stars were retained as targets. In addition, a domain of subgiant and giant stars was constructed, as shown in the rectangle at the upper right of Figure \ref{fig2}. These 100 nearest evolved stars were retained as targets. This selection of main sequence, subgiant, and giant stars between 5-50 pc yields 1649 target stars. 

\subsection{The Combined Stellar Sample}

\begin{figure}
\epsscale{.90}
\plotone{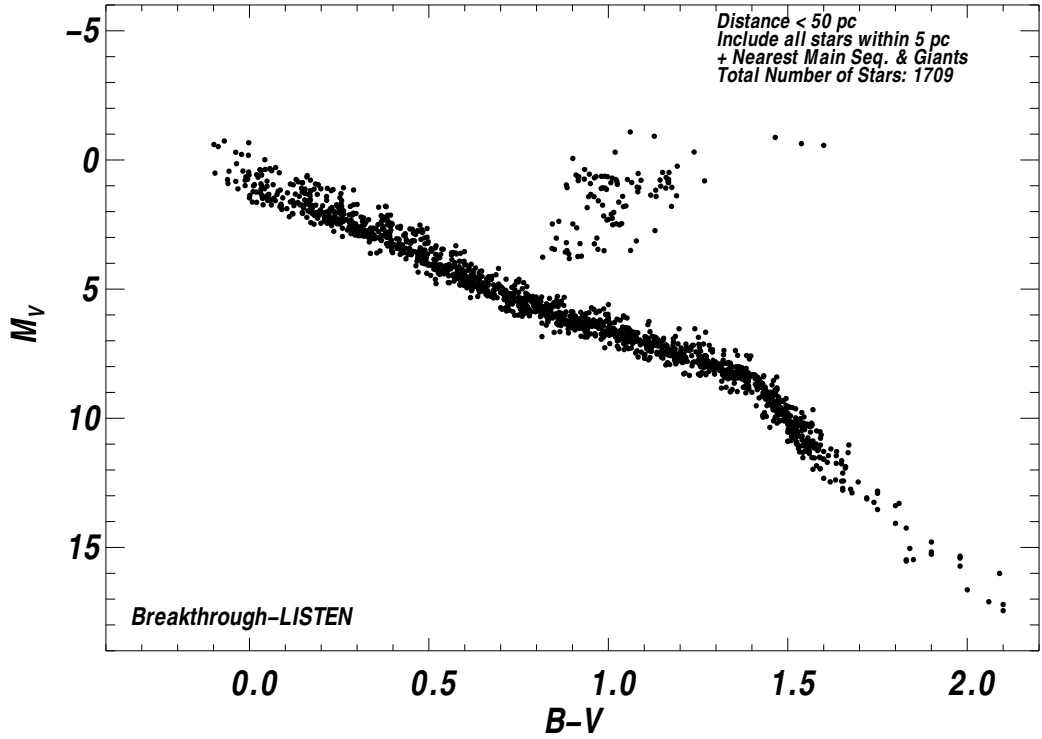}
\caption{The H-R Diagram of the 5-50 pc and 5 pc samples of target stars for \BLp. Domains of $B-V$ and $M_V$ were constructed along the Main Sequence, within which the nearest 100 stars were selected to be included in the final \BL target list. In addition, a domain of giant stars was constructed and the nearest 100 such stars were identified. 
}
\label{fig3}
\end{figure}

 The H-R Diagram of the combined sample of 1709 stars is shown in Figure \ref{fig3} and listed in Table \ref{table:stars_table}.  Inclusion of stars across the main sequence suppresses bias against any type of star. The nearest stars, within 5 pc, are composed mostly of M dwarfs, down to spectral type M8 $(B-V>2)$, near the brown dwarf boundary. We retained stars having hydrogen-burning ages of less than a few billion years (stars more than twice the mass of the Sun) despite the concern about adequate time for evolution to yield intelligence, as well as giants, which evolved off the main sequence less than 100 million years ago. 

\begin{figure*}
\epsscale{0.9}
\plotone{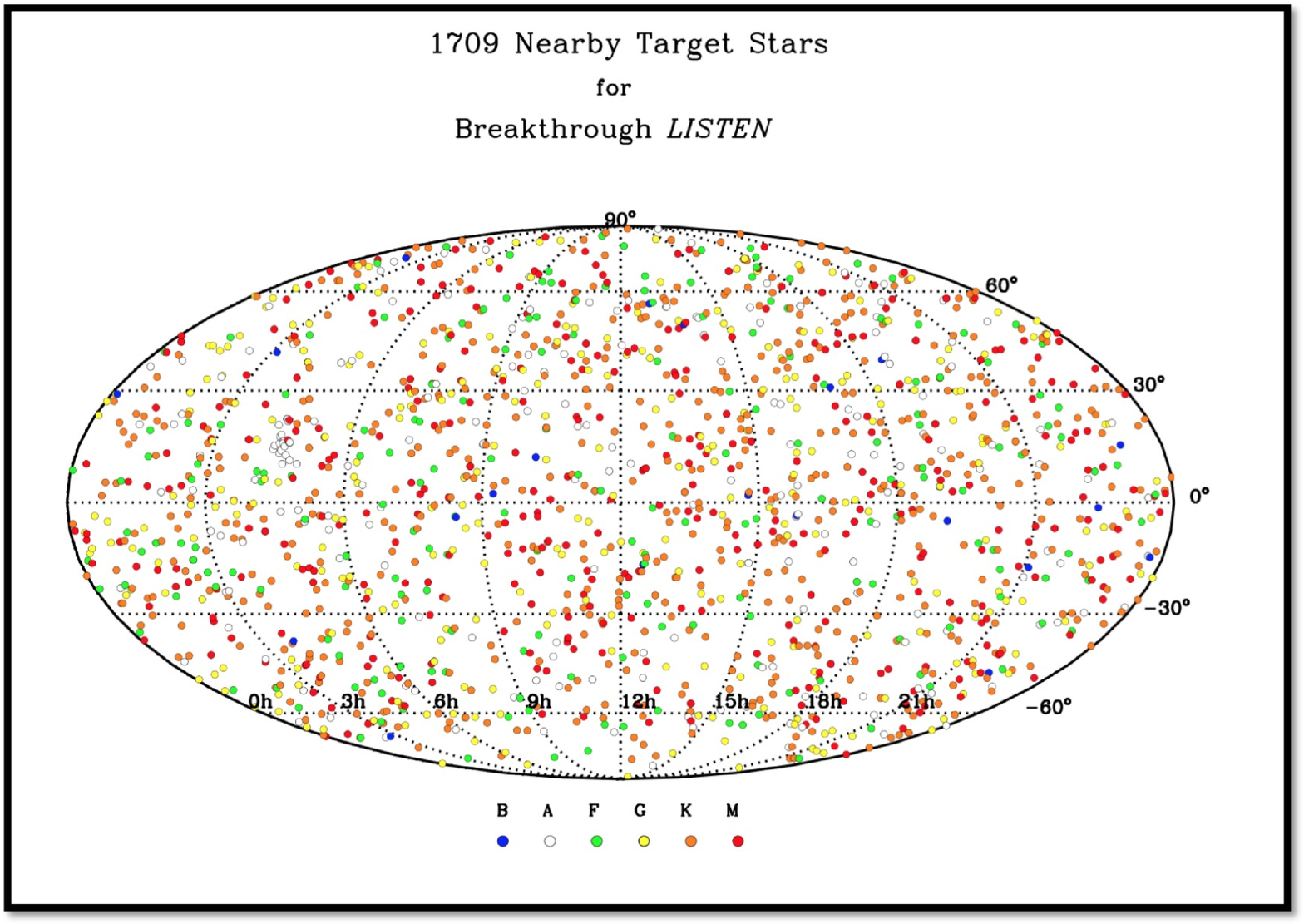}
\caption{The equatorial coordinates of the entire set of \BL target stars. It includes all stars within 5 pcs and the nearest stars within B-V domains in a 5-50 pc sub-sample. The goal is to construct a sample of target stars for \BL that has a sampling of the diversity of stars found within 50 pc.
}\label{fig4}
\end{figure*}

This survey spreads the investment of telescope time across a broad distribution of the nearest main sequence, sub-giant, and giant stars, with special inclusion of all stars within 5 pc. Figure \ref{fig4} shows the position of all 1709 targets (main sequence and giant stars) on the sky in equatorial coordinates. Each star is color coded to represent its spectral type (BAFGKM), from hottest to coolest surface temperature.

\begin{figure}
\epsscale{.90}
\plotone{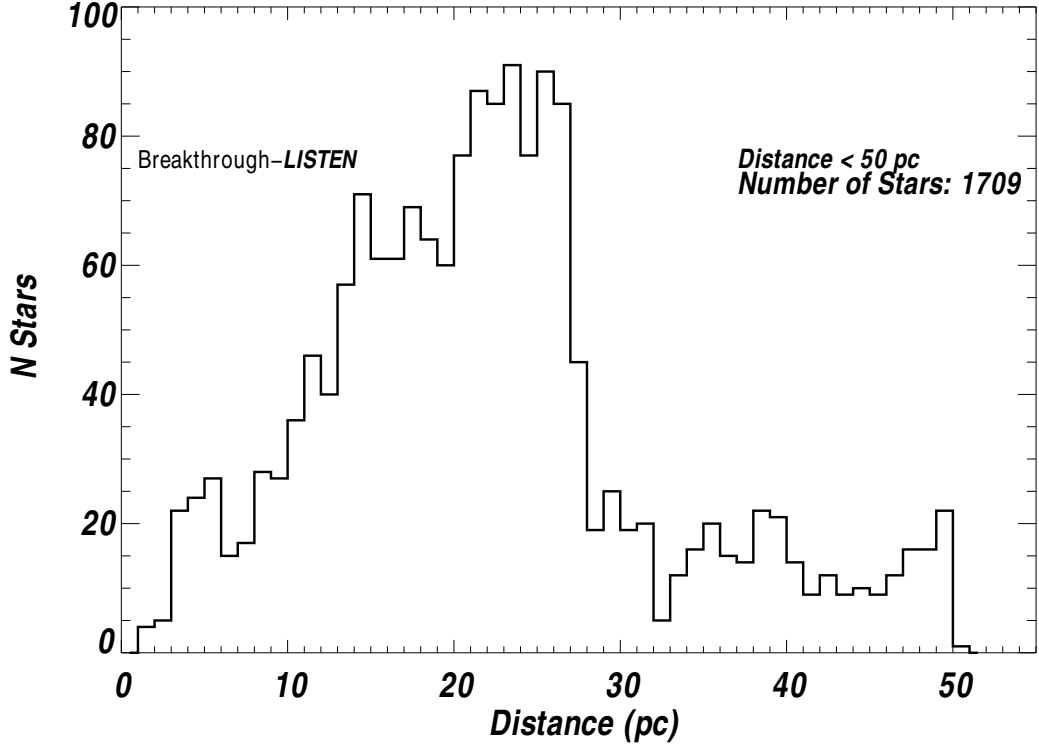}
\caption{The histogram of distances to all 1709 target stars of the \BL survey. The nearest stars (at far left) are typically M dwarfs and the bulk of the stars from 10-30 pc are typically FGKM stars. The stars beyond 30 pc are F, A, and B stars and giants, for which their low number density in the Galaxy necessitates greater distances to include the nearest of them.
}\label{fig5}
\end{figure}

The histogram of distances to the entire target sample of 1709 stars is shown in Figure \ref{fig5}. The distances range from 1.3 pc (the Alpha Cen triple system) to 50 pc. The peak is near 25 pc, representing the typical distance to the nearest 100 FGK stars within 50 pc. The M dwarfs are concentrated at distances under 20 pc, due to their high number density in the Galactic disk and to the purposeful inclusion of all stars within 5 pc that are mostly M dwarfs. The most distant stars in the sample, located beyond 30 pc, are composed mostly of early G, F, A, and B type stars and giants that have a lower number density for which the nearest are farther away. 

\section{Galaxies}

As SETI targets, galaxies offer both challenges and opportunities.  The great distances to the nearest large galaxies, measured in millions of light years, require extraterrestrial technologies to generate luminosities of order 10$^7$ TW of equivalent isotropically radiated power (EIRP) to be detected, i.e. a million times the EIRP of the most powerful transmitter at Arecibo.  Still, this power requirement is only a technological hurdle, not an obstacle of physical principle.  The angular sizes of nearby galaxies are typically tens of arcminutes, comparable to the beam sizes of the Green Bank and Parkes telescopes, depending on frequency and dish size.  Thus, a single radio observation captures any SETI signals coming from anywhere within a large fraction of an entire galaxy, including billions of stars and their planets, simultaneously.  If advanced civilizations are very rare, but occasionally produce energetic radio emission of over 10$^7$ TW, such radio observations of galaxies would sample that domain of the ETI luminosity function. Indeed, few SETI programs have included galaxies in their target list, making them fresh hunting ground. 

The selection process of target galaxies for the \BL program spans the entire sky and includes those galaxies to be the nearest representatives of the five major morphological classes of galaxies. With all due appreciation of the habitability of our home spiral galaxy, we find no compelling argument that technological life might preferentially arise in one type of galaxy over any other, as long as the abundance of heavy elements is within an order of magnitude of solar. Specifically, we have included 40 spirals, 40 ellipticals, 20 dwarf spheroidals, 20 irregulars, and three S0 galaxies. We constructed the sample from two sources: NEARGALCAT, a catalog of all known galaxies within 11 Mpc \citep{Karachentsev2013} and \cite{Djorgovski1987} which contains properties of 85 nearby, regular giant elliptical galaxies.

The NEARGALCAT contains all 869 known galaxies within 11 Mpc, including spirals, dwarf spheroidals, and irregulars. However, NEARGALCAT contains only three normal giant elliptical galaxies, reflecting their general paucity in the Universe. Indeed, \cite{Ann2015} show that normal giant elliptical galaxies comprise only 1.5\% of all galaxies. In contrast, spirals make up 32\% of all galaxies and irregulars contribute 42\%. Normal elliptical galaxies occur mostly in galaxy clusters, such as Coma or Virgo.

To include normal giant ellipticals in the \BL target list requires that we reach out to 30 Mpc. From \cite {Djorgovski1987}, which established the Fundamental Plane, we identify a subset of ellipticals having $z < 0.01$, systematically the closest ellipticals from that paper which included galaxies at $z < 0.025$. The galaxies in \cite{Djorgovski1987} are biased toward the northern hemisphere, all having a declination $ > -25$ deg. However, this bias does not matter for \BLp, as over half of these galaxies are accessible by both GBT and Parkes, allowing a free allocation of targets to each. 

\begin{figure}
\epsscale{.90}
\plotone{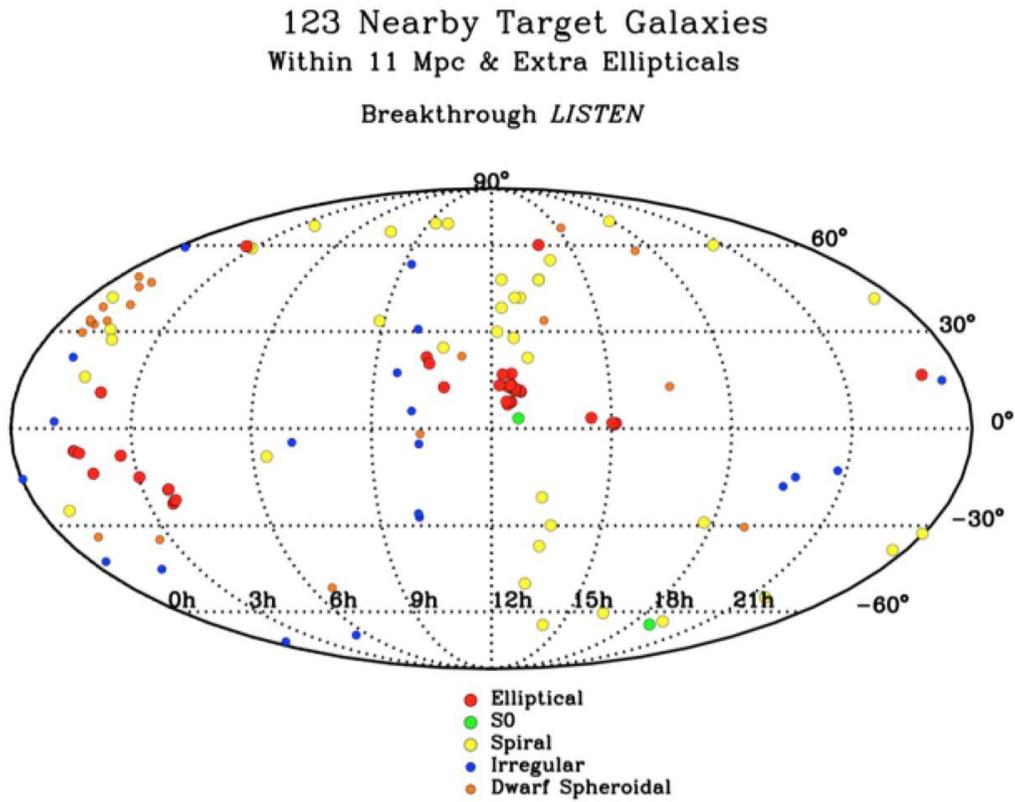}
\caption{A map, in equatorial coordinates, of the \BL target galaxies. This spatial distribution is non-uniform because of natural clumping and association rather than selection effects.
}
\label{fig10}
\end{figure}

Analogous to the 5-50 pc sample of stars across all spectral types described above, we chose a roughly uniform distribution of the different types of the nearest galaxies in all five morphological classes: 40 nearest spirals, 40 nearest ellipticals, 20 nearest irregulars, 20 nearest dwarf spheroidals, and three nearby S0 galaxies (Table \ref{table:gal_table}). Figure \ref{fig10} shows the distribution of the selected galaxies on the celestial sphere. The angular sizes of most of these 123 nearby galaxies are less than 10 arcminutes. Of course the very nearest galaxies, such as Andromeda, have angular sizes over 30 arcminutes, depending on which isophot one adopts as the edge. Observations of galaxies with GBT and Parkes may consist of one, or at most a few, pointings near the center of each galaxy. The typical radio beam size at most frequencies will cover a large fraction of the stars in the nuclei, bulges, and outskirts of most of these galaxies.

We will observe the galactic plane and bulge of the Milky Way over 1200-1550 MHz using the 21-cm multibeam receiver on the Parkes 64-m radio telescope \citep{multibeam1996}. This broad Milky Way survey will include billions of the stars located within the densest regions of our galaxy. The Parkes telescope is well suited to this task, due to the southern declination of the Galactic center and the high survey speed afforded by the 21-cm multibeam receiver. We will employ a step-and-stare mosaic strategy, as used in the High Time Resolution Universe survey \citep{keith2010}, to survey roughly 3000 square degrees over galactic longitudes
 $-174\deg < l < 60\deg$ and latitudes $|b| < 6.5\deg$. To leverage potential scintillation-induced signal amplification and intermittency \citep{cordes1997}, each sky position will be observed several times. We anticipate observing the ~3000 square degree survey area over 1500 hours total, leading to roughly 1080 s total dwell time per pointing. 
 
For high northern latitude portions of the galactic plane, the Parkes observations will be supplemented with appropriate parity using the GBT.


The two nearest galaxies outside of the Milky Way, the Large and Small Magellanic Clouds (LMC, SMC) serve as galactic targets that are 10 times closer than M31. The Parkes Telescope has previously been used to survey the Magellanic Clouds for continuum sources \citep{Haynes1991} and for SETI signals \citep{Shostak1996}. The latter SETI survey searched 3 pointings (1/1000 of the projected angular size of the SMC) over the bandpass from 1.2 - 1.75 GHz using the Project Phoenix hardware but found no narrow band or slow pulsed signals greater than 19 Jy.

Due to its large angular size on the sky, roughly 100 square degrees, LMC and SMC are target galaxies with over 10$^{10}$ and 10$^{7}$ stars, respectively \citep{Vaucouleurs1991}. Using the 21 cm multi-beam receiver with a similar step and stare strategy to that proposed for the Galactic Plane Survey, with a dwell time of 1080 s dwell time per pointing, the two surveys could be conducted in 50 hours and 10 hours of observing time, respectively.



\section{Exotica}
In addition to stars and galaxies, the \BL target list will also include an “exotica” category.  This category contains classes of astronomical objects that seem less likely to harbor technological life as we expect it, e.g. environments very different than our own planet, but present intriguing opportunities for more speculative investigations.  Fundamentally, these objects merit SETI observations in case our anthropocentric expectations are wrong. Such a list will include, but not be limited to: brown dwarfs, white dwarfs, pulsars, black holes, asteroids, KBOs, Pluto, Sedna, the Moon, pulsars, and active galactic nuclei such as the M87 nucleus. Solar system asteroids will also be considered for observation as suggested by \cite{Gertz2016}. We expect less than 5\% of observing time will be spent on exotic objects.

\section{Observing Strategy}

When observing the 5 pc sample, the 5-50 pc sample and the galaxy sample, we implement an on-source/off-source strategy. Each primary target is observed three times with an observation of an off source target observed after each primary target observation. In all cases, we move the telescope off of the primary target to ensure that any emission from the primary target is sufficiently attenuated in off-source observations. We initially observed the off-source in a uniform way by moving, one degree in declination away from the primary source. We then modified our strategy to observe known stars as off-source targets. This set of off-source stars is composed of the brightest 40,000 stars from the Hipparcos catalog that are not in already in the primary sample. Having such a large number of stars ensures that we do not need to spend excessive telescope time on slew times from the primary (A) to off-source targets (B, C, D). For each primary star, we choose three nearby off-source stars and observe in the pattern ABACAD. So that the signal to noise ratio is roughly the same, we use the same exposure time for each pointing. 

Special care is required when observing galaxies that are extended on the sky. The same on/off source  strategy is employed, with slight modifications to account for the angular extent of the galaxy and to ensure optimal on-source time. A single pointing allows us to observe the majority of stars within a galaxy, and with the off source positioning, we can distinguish local radio interference from signals originating in the observed galaxy.

\section{Observations to Date}
\label{obs_to_date}
\subsection{Green Bank Telescope}
 In 2016, we observed the 5 pc and 5-50 pc samples at L-band for stars above declination of $-20$ degrees. L-band ($1.1-1.9$ GHz) was chosen as the first receiver as it is a well characterized receiver and well matched to early incarnations of the \BL digital instrumentation. We are now moving to S-band, spanning $1.7 - 2.6$ GHz, and later will extend to higher frequencies as we gain confidence with our instrumentation and install additional processing capacity.


 The standard cadence for a single target is a set of six, five-minute observations ABACAD as described above. From each five minute observation, we compute three separate data products, the sum of which is roughly 2 percent of the volume of the raw voltage data. A Graphics Processing Unit (GPU) accelerated code is used to produce three dynamic spectra data files with resolutions of 3 Hz, 366 Hz, and 3 KHz. The volume of each 5 minute observation is 1 TB per data writing disk. In L-band, the bandwidth is 800 MHz, requiring 4 computers to simultaneously write to disk. With the full future capacity of 10 GHz of bandwidth, 64 computers will write data to disk at the combined rate of 64 TB of raw voltage data per 5 min observation.  The final data product will be 1.28 TB of data per 5 minute observation of dynamic spectra data. Similar data products will be produced for data acquired with the Parkes Telescope. As much data as possible will be made publicly available\footnote{\url{http://www.breakthroughinitiatives.org/OpenDataSearch}}.

\subsubsection{Data Quality Analysis, Raw Voltage}

While the primary archival data product is total power dynamic spectra, the raw time-domain (voltage) data used to produce dynamic spectra remain temporarily available in order to conduct focused signal analysis. These data enable the search for wider bandwidth signals through detection of communication features such as circularity, cyclostationarity or higher-order statistics. They are also essential to compensate coherently for Doppler frequency shifts induced by relative accelerations (e.g. correcting to ``magic'' reference frames), identifying subtle terrestrial radio frequency interference, and more broadly, monitoring the system and data quality through statistical analysis. A raw data software package is being developed to provide to any user the tools to perform signal processing, analysis and detection on raw voltages that may become available in the future.

\subsubsection{Data Quality Analysis, W3 Maser}
\begin{figure}
\epsscale{.90}
\plotone{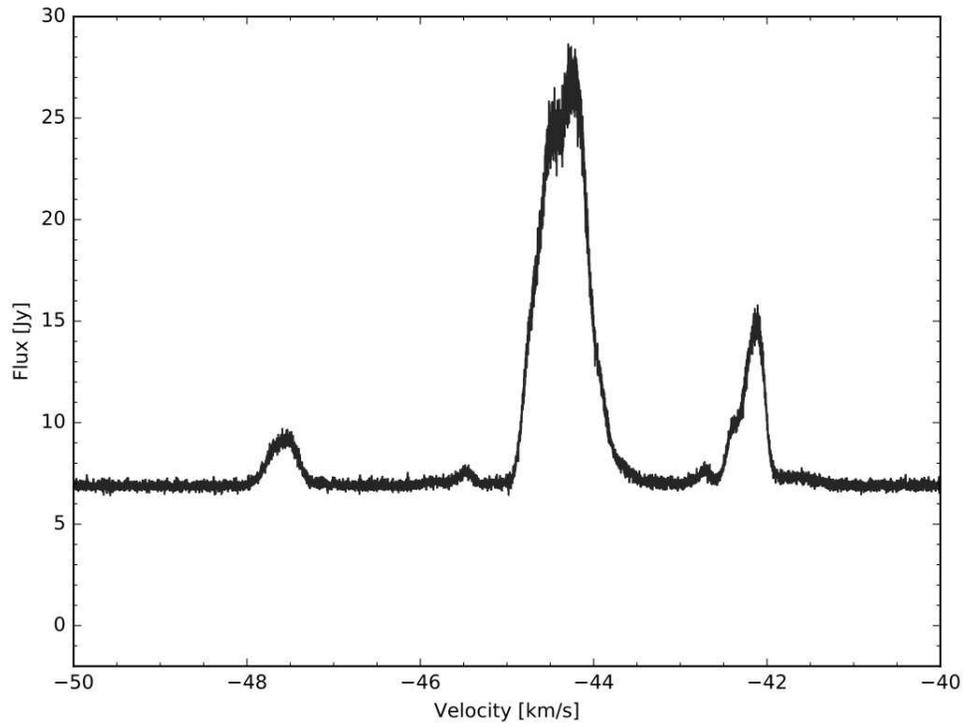}
\caption{Example \BL data showing Stokes I spectra of OH maser emission in the W3 star-forming region. The plotted data correspond to 0.2 MHz, centered at 1667.5 MHz, with 2.8 Hz channel resolution. 
}
\label{w3plot}
\end{figure}

In early tests of our observing system, we observed the W3(OH) star-forming region using the L-band receiver of the GBT with the BL data recorder systems (MacMahon, in prep). Figure \ref{w3plot}  shows the Stokes I spectra of OH maser emission in the W3 star-forming region, generated from BL observations on MJD 57424.92. The figure shows an extracted 0.2 MHz band, centered at 1667.5 MHz, with 2.8 Hz channel resolution, using a 20 second integration period.  See \cite{WrightM2004} for comparison.

\subsubsection{Data Quality Analysis, Voyager 1 Observations}
\begin{figure}
\epsscale{.90}
\plotone{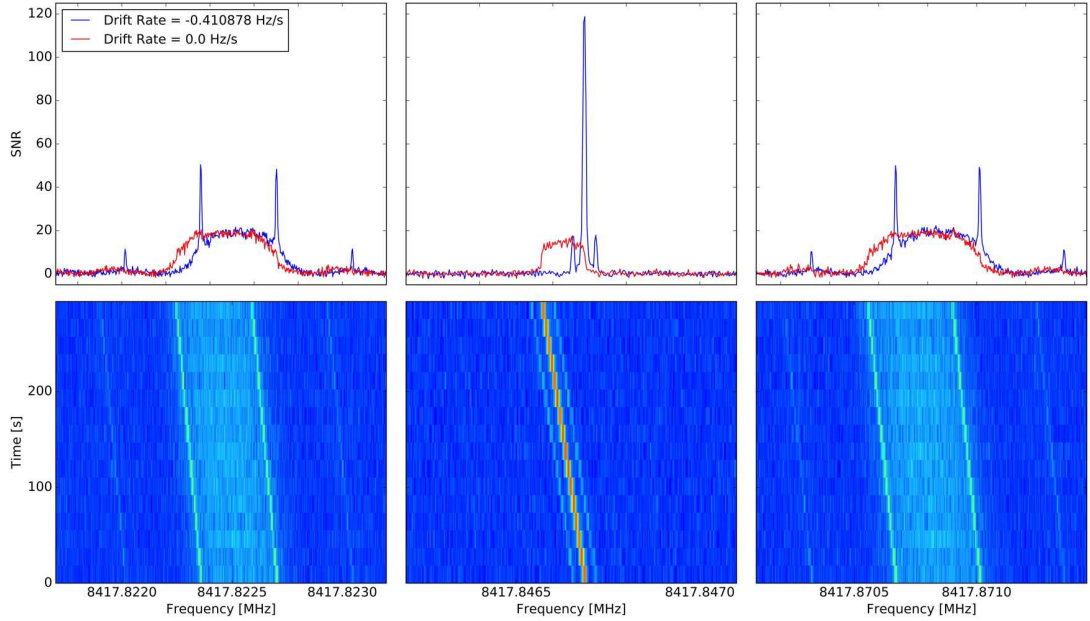}
\caption{A 5 min long observation of the Voyager 1 spacecraft using the GBT showing the Voyager I carrier wave and drift pattern. The top plots show the integrated spectrum with and without drift rate correction. The bottom plots show the drifting signals with time. The coarseness of the 18 sec time integration of the filterbank is visible in the pixelation of the time axis. Each time bin is not drift-corrected, limiting our search to incoherent drift corrections. Motion of the Earth relative to Voyager 1 causes received signals to appear near 8417.8 MHz, offset from the native downlink frequency of Voyager 1 at 8415 MHz.
}
\label{voyager}
\end{figure}

Narrow band signals have been the favorite among ETI searches for a variety of reasons, but especially due to their obviously artificial nature.  In particular, most searches look for narrow band signals that drift in frequency over time. Drifting signals are produced when a Doppler acceleration exists between the transmitter and the receiver, typically signals that are not stationary at the Earth's surface. However, it is still possible to have signals that are man-made and also far away from the earth, such as the Voyager 1 spacecraft. Its signal drifts in frequency due to the relative acceleration between the Earth and the spacecraft in an uncorrected frame of reference. Figure \ref{voyager} shows a five minute observation with the GBT of the Voyager 1 spacecraft, the carrier signal as well as the two side bands. The lower plots show the dynamic spectrum of each of the three signals. The plots on the top show the integrated spectrum with and without the drift correction at the best fit drift rate. This serves as a proof of concept on how the drift rate search is needed to detect an extra terrestrial signal, man-made or not.

Observations like the Voyager signal are a good test of our search algorithms due to their drift and narrow band properties. We analyze narrow-band drifting signals for a wide range of drift rates (eg. $\textit{dr = } \pm$ 10 Hz $s^{-1}$). The 18 sec time bins together with the 3 Hz frequency resolution allow for a change in drift rate, $\Delta \textit{dr}$, of 0.167 Hz $s^{-1}$. This is sufficient to have a good characterization of a signal wider than the frequency resolution. Additional details on this analysis pipeline and presentation of results of early observations are forthcoming (Enriquez in prep.).

\subsection{Automated Planet Finder}

The APF is observing the 5 pc and 5-50 pc spectral type complete sample. The limiting magnitude for pointing on the APF is $V=14$. We choose to observe only targets brighter than $V=12.0$ in order to acquire sufficient throughput in the 20 minute observations. Due to the similar optical qualities and wavelength coverage between the Levy Spectrometer and the HIRES spectrometer on Keck, we choose to exclude stars already observed with Keck/HIRES by the California Planet Search (CPS). All spectra observed with Keck/HIRES are available to the public online, once the initial proprietary period is past( $12-18$ months). Since the CPS targets are slowly rotating FGK stars, the 5-50 parsec sample that will be observed with the APF includes many young stars, active stars, rapid rotators and stars hotter than 6200 K, which are typically excluded from radial velocity planet searches.  Our sensitivity to laser lines will match that of \cite{Tellis2015} to the order of 1 photon per m$^-$$^2$ s$^-$$^1$. Extinction due to interstellar dust will limit detection of laser lines to a few thousand light years from the Earth.

Of the 1185 stars above a declination of $-20$ degrees, 414 of them have been previously observed with Keck/HIRES. Of the remaining 771, 560 of them have been observed since the \BL APF program was initiated. All of the spectra are available for download from the \BL public archive\footnote{\url{http://www.breakthroughinitiatives.org/OpenDataSearch}}.

\section{Conclusions}
In an effort to conduct the most complete and unbiased SETI search possible, we choose targets for a focused SETI search that includes stars, galaxies, and other exotic targets including brown dwarfs and asteroids. Using new and established instruments at the Green Bank Observatory, Parkes Telescope and APF, we are undertaking targeted searches in the radio and at optical wavelengths, with radio telescopes that together cover the entire sky.  

Our goal remains to make a detection of a candidate extraterrestrial signal, amenable to follow-up observations that provide validation (or rejection) and characterization.  If we are left with merely non-detections, we must provide quantitative upper limits on the flux density at all observed frequencies and for all targets, with the goal of defining the science goals for the next generation of SETI searches, ensuring they sample distinct domains of parameter space of ETI signals. Indeed we see promise with current and future radio arrays, and with next-generation optical/IR surveys of the entire sky sampled at high cadence.



\acknowledgements{Acknowledgments: We thank Gloria and Ken Levy for support of the Automated Planet Finder Spectrometer. We gratefully thank Jill Tarter, Seth Shostak, Frank Drake, Jason Wright, John Gertz, and Andrew Fraknoi for many conversations about the search for intelligent life in the universe. This work made use of the SIMBAD database (operated at CDS, Strasbourg, France) and NASA’s Astrophysics Data System Bibliographic Services. Funding for \BL research is sponsored by the Breakthrough Prize Foundation.}


{\it {Facilities: Green Bank Telescope, Automated Planet Finder, Lick Observatory, Parkes Observatory}}

\clearpage



\clearpage

\begin{deluxetable}{lcccccccc}
\tabletypesize{\footnotesize}
\tablecaption{Stellar Targets}
\tablewidth{0pt}
\tablehead{
  \colhead{Star Name} & \colhead{RA}  & 
  \colhead{Dec}      & \colhead{Epoch}  & \colhead{V mag} & 
  \colhead{Sp Type}& \colhead{Distance (pc)} & \colhead{PM RA}  & \colhead{PM Dec} 
}
\startdata
       HIP2 &  00:00:01.0 &   -19:29:55 &    2000 &    9.27 &     K3V & 45.6  &    0.18 &    0.00 \\
     HIP169 &  00:02:08.7 &   -68:16:50 &    2000 &    9.24 &     M0V & 15.8  &    0.21 &   -0.23 \\
     HIP171 &  00:02:10.2 &    27:04:56 &    2000 &    5.80 &     G3V & 12.4  &    0.78 &   -0.92 \\
     HIP194 &  00:02:29.7 &    08:29:07 &    2000 &    5.70 &     F0V & 37.6  &   -0.09 &   -0.05 \\
     HIP400 &  00:04:56.3 &    23:16:10 &    2000 &    7.82 &     G9V & 25.6  &    0.38 &   -0.01 \\
     HIP428 &  00:05:10.9 &    45:47:11 &    2000 &    9.95 &      M2 & 11.4  &    0.87 &   -0.15 \\
     HIP436 &  00:05:17.7 &   -67:49:57 &    2000 &    8.49 &     K5V & 16.0  &   -0.12 &   -0.56 \\
     HIP443 &  00:05:20.1 &   -05:42:27 &    2000 &    4.61 &   K1III & 39.4  &   -0.01 &    0.09 \\
\enddata     
\label{table:stars_table}
\end{deluxetable}

\clearpage

\begin{deluxetable}{lcccccc}
\tabletypesize{\footnotesize}
\tablecaption{Galaxies}
\tablewidth{0pt}
\tablehead{
  \colhead{Galaxy} & \colhead{RA}  & 
  \colhead{Dec}      & \colhead{B mag}  & \colhead{K mag} & 
  \colhead{Distance (Mpc)} & \colhead{Morphology}
}
\startdata
Milky Way  &17:45:40.0 & -29:00:28 & -4.6 & 0.0 & 0.01 & Spiral  \\
MESSIER031  &00:42:44.6 & 41:16:09 & 4.2 & 0.9 & 0.77 & Spiral  \\
MESSIER033  &01:33:50.8 & 30:39:37 & 6.2 & 4.1 & 0.85 & Spiral  \\
Dw1  &02:56:56.0 & 58:54:42 & 15.0 & 5.1 & 2.80 & Spiral  \\
Maffei2  &02:41:54.6 & 59:36:11 & 14.7 & 4.5 & 2.80 & Spiral  \\
NGC2403  &07:36:51.5 & 65:35:57 & 8.8 & 6.1 & 3.18 & Spiral  \\
IC0342  &03:46:48.7 & 68:05:44 & 9.3 & 4.5 & 3.28 & Spiral  \\
NGC5102  &13:21:58.0 & -36:37:46 & 10.2 & 6.9 & 3.40 & Spiral  \\
\enddata     
\label{table:gal_table}
\end{deluxetable}

\clearpage

\bibliographystyle{apj}
\bibliography{targets}

\end{document}